\documentclass
	[
		10 pt
	]
	{article}
\usepackage
	[
		a4paper,
		left = 2.500 cm,
		right = 2.500 cm,
		top = 2.500 cm,
		bottom = 3.000 cm
	]
	{geometry}
\usepackage{tikz, pgf, pgfplots}
\usetikzlibrary
	{
		3d,
		arrows,											
		arrows.meta,
		automata,										
		backgrounds,
		bending,
		calc,
		chains,
		datavisualization.formats.functions,			
		decorations.pathmorphing,
		decorations.pathreplacing,
		decorations.text,
		fadings,
		fit,
		graphs,
		graphs.standard,
		matrix,
		patterns,
		positioning,									
		quantikz2,
		quotes,
		shadings,
		shadows,
		shadows.blur,
		shapes,
		shapes.geometric,
		trees
	}
\pgfplotsset{compat = newest}
\usepgflibrary
	{
		shapes.misc
	}
\usepgfplotslibrary
	{
		dateplot
	}
\usepackage{tikzpeople}
\usepackage{fontawesome6}
\usepackage{smartdiagram}

\allowdisplaybreaks
\usepackage{amsthm}
\usepackage{amssymb}
\usepackage[bb = boondox]{mathalfa}
\usepackage{dsfont}
\usepackage{newtxmath}
\usepackage{mathtools}
\usepackage{multicol}
\usepackage{physics2}
\usephysicsmodule{ab.legacy}							
\usephysicsmodule{braket}								
\usephysicsmodule{diagmat}								
\usephysicsmodule{nabla.legacy}							
\usepackage{empheq}
\usepackage[ compat = newest ]{yquant}

\usepackage{sectsty}
\allsectionsfont{\color{WordBlue}}
\usepackage[x11names]{xcolor}
\usepackage{varwidth}
\usepackage{graphicx}
\usepackage{wrapfig}
\usepackage{subcaption}
\usepackage{xurl}                                                                                  %
\usepackage{hyperref}
\hypersetup
	{
		colorlinks,
		linkcolor = {RedPurple},
		citecolor = {WordBlueDarker25},
		urlcolor = {WordAquaDarker50}
	}
\usepackage{tabularray}
\usepackage{booktabs}
\usepackage{colortbl}
\usepackage{float}
\usepackage{nicematrix}
\usepackage{cellspace}
\usepackage{makecell}
\usepackage{multirow}
\usepackage{caption}
\usepackage[ linesnumbered, tworuled, vlined ] {algorithm2e}
\usepackage{appendix}
\usepackage{enumitem}
\usepackage{siunitx}
\usepackage{xintexpr}
\usepackage{spreadtab}
\usepackage{framed}
\usepackage{svg}
\usepackage{lipsum}
\usepackage{listings}
\definecolor{MyLightRed}{RGB}{244, 213, 245}
\definecolor{Purple}{HTML}{911146}
\definecolor{PurpleDark}{RGB}{102, 0, 102}
\definecolor{RedDarkLight}{HTML}{ea005f}
\definecolor{RedDarkLightest}{HTML}{ff0088}
\definecolor{RedPurple}{HTML}{AA007F}
\definecolor{WordPinkAccent1Darker25}{HTML}{B3186D}
\definecolor{WordPinkAccent1Darker50}{HTML}{781049}
\definecolor{WordPinkAccent1Lighter40}{HTML}{EE80BC}
\definecolor{WordPinkAccent1Lighter60}{HTML}{F3AAD2}
\definecolor{WordPinkAccent1Lighter80}{HTML}{F9D4E8}
\definecolor{WordRed}{RGB}{255, 0, 102}
\definecolor{WordRedAccent5Lighter60}{HTML}{F5B5A7}
\definecolor{WordRedAccent5Darker25}{HTML}{B23214}
\definecolor{GreenDark}{HTML}{225522}
\definecolor{GreenLighter1}{HTML}{00B383}
\definecolor{GreenLighter2}{HTML}{00AA7F}
\definecolor{GreenLightest}{HTML}{00FFA0}
\definecolor{GreenTeal}{HTML}{008080}
\definecolor{WordLightGreen}{RGB}{140, 214, 192}
\definecolor{WordGreen}{RGB}{0, 176, 80}
\definecolor{BlueVeryDark}{HTML}{222255}
\definecolor{MyBlue}{RGB}{0, 64, 128}
\definecolor{MyDarkBlue}{RGB}{0, 51, 102}
\definecolor{MyVeryLightBlue}{RGB}{211, 245, 247}
\definecolor{WordBlue}{RGB}{19, 65, 99}
\definecolor{WordBlueDark}{RGB}{46, 116, 181}
\definecolor{WordBlueDarker}{RGB}{31, 78, 121}
\definecolor{WordBlueDarker25}{RGB}{54, 96, 146}
\definecolor{WordBlueDarker50}{RGB}{36, 64, 98}
\definecolor{WordBlueDarkest}{RGB}{0, 32, 96}
\definecolor{WordBlueLight}{RGB}{0, 112, 192}
\definecolor{WordBlueVeryLight}{HTML}{00B0F0}
\definecolor{WordIceBlue}{RGB}{223, 227, 229}
\definecolor{MagentaDark}{RGB}{106, 65, 152}
\definecolor{MagentaLight}{RGB}{128, 100, 162}
\definecolor{MagentaLighter}{RGB}{161, 106, 221}
\definecolor{MagentaVeryDark}{RGB}{97, 75, 128}
\definecolor{MagentaVeryLight}{RGB}{178, 162, 201}
\definecolor{WordAquaAccent1Darker25}{HTML}{276E8B}
\definecolor{WordAquaAccent1Darker50}{HTML}{1A495D}
\definecolor{WordAquaAccent1Lighter40}{HTML}{7FC0DB}
\definecolor{WordAquaAccent1Lighter60}{HTML}{A9D5E7}
\definecolor{WordAquaAccent1Lighter80}{HTML}{D4EAF3}
\definecolor{WordAquaAccent2Darker25}{HTML}{398E98}
\definecolor{WordAquaAccent2Darker50}{HTML}{265F65}
\definecolor{WordAquaAccent2Lighter40}{HTML}{9AD3D9}
\definecolor{WordAquaAccent2Lighter60}{HTML}{BCE1E5}
\definecolor{WordAquaAccent2Lighter80}{HTML}{DDF0F2}
\definecolor{WordAquaDarker25}{HTML}{31869B}
\definecolor{WordAquaDarker50}{HTML}{215967}
\definecolor{WordAquaLighter40}{HTML}{92CDDC}
\definecolor{WordAquaLighter60}{HTML}{B7DEE8}
\definecolor{WordAquaLighter80}{HTML}{DAEEF3}
\definecolor{WordDarkerTeal}{RGB}{48, 82, 80}
\definecolor{WordDarkTeal}{RGB}{72, 123, 119}
\definecolor{WordDarkTealLighter80}{RGB}{207, 223, 234}
\definecolor{WordLightTeal}{RGB}{160, 199, 197}
\definecolor{WordVeryLightTeal}{RGB}{223, 236, 235}
\definecolor{WordTurquoiseLighter80}{RGB}{209, 238, 249}
\definecolor{Brown}{HTML}{666633}
\definecolor{WordGoldAccent1Darker25}{HTML}{C49A00}
\definecolor{WordGoldAccent1Lighter40}{HTML}{FFDF6A}
\definecolor{WordOrangeAccent2Lighter60}{HTML}{FCD3A4}
\definecolor{WordOrangeAccent4Lighter60}{HTML}{F7C5A1}
\definecolor{LavenderBlush}{RGB}{255, 240, 245}
\definecolor{MediumTurquoise}{RGB}{72, 209, 204}
\definecolor{PowderBlue}{RGB}{176, 224, 230}
\definecolor{SkyBlue}{RGB}{135, 206, 235}
\definecolor{Azure2}{RGB}{224, 238, 238}
\definecolor{Azure3}{RGB}{193, 205, 205}
\definecolor{CadetBlue4}{RGB}{83, 134, 139}
\definecolor{DarkSeaGreen1}{RGB}{193, 255, 193}
\definecolor{DeepPink4}{RGB}{139, 10, 80}
\definecolor{Honeydew2}{RGB}{224, 238, 224}
\definecolor{LightSkyBlue1}{RGB}{176, 226, 255}
\definecolor{LightSkyBlue3}{RGB}{141, 182, 205}
\definecolor{LightSkyBlue4}{RGB}{96, 123, 139}
\definecolor{LightSteelBlue1}{RGB}{202, 225, 255}
\definecolor{LightSteelBlue4}{RGB}{110, 123, 139}
\definecolor{MediumPurple1}{RGB}{171, 130, 255}
\definecolor{PaleTurquoise3}{RGB}{150, 205, 205}
\definecolor{PaleVioletRed3}{RGB}{205, 104, 137}
\definecolor{Purple1}{RGB}{155, 48, 255}
\definecolor{SeaGreen1}{RGB}{84, 255, 159}
\definecolor{SeaGreen2}{RGB}{78, 238, 148}
\definecolor{SeaGreen3}{RGB}{67, 205, 128}
\definecolor{SkyBlue1}{HTML}{87CEFF}
\definecolor{SkyBlue4}{RGB}{74, 112, 139}
\definecolor{SteelBlue1}{RGB}{99, 184, 255}
\definecolor{Thistle3}{RGB}{205, 181, 205}
\definecolor{Turquoise4}{RGB}{0, 134, 139}
\definecolor{VioletRed1}{RGB}{255, 62, 150}
\definecolor{VioletRed2}{RGB}{208, 32, 144}
\definecolor{VioletRed3}{RGB}{199, 21, 133}
\definecolor{VioletRed4}{RGB}{139, 10, 80}
\usepackage
	[
		most
	]
	{tcolorbox}
\newcounter{MyCorollary}[section]
\renewcommand{\theMyCorollary}{\thesection.\arabic{MyCorollary}}
\newtcolorbox{corollary} [ 1 ] [ ]
	{
		breakable,
		enhanced,
		enhanced jigsaw,
		skin = enhanced,
		attach boxed title to top left = { xshift = -5.000 mm, yshift = 0.000 mm },
		boxed title style = { boxrule = 0.000 pt, sharp corners = all },
		colbacktitle = RedPurple!90!black,
		coltitle = white,
		fonttitle = \bfseries,
		varwidth boxed title,
		colback = RedPurple!03,
		colframe = RedPurple,
		sharp corners = all,
		toprule = 0.000 mm,
		bottomrule = 0.500 mm,
		leftrule = 0.500 mm,
		rightrule = 0.000 mm,
		code = { \refstepcounter{MyCorollary} },
		title = {Corollary~\theMyCorollary:\if\relax\detokenize{#1}\relax\else~#1\fi},
	}
\newtcolorbox{corollary*} [ 1 ] [ ]
	{
		breakable,
		enhanced,
		enhanced jigsaw,
		skin = enhanced,
		attach boxed title to top left = { xshift = -5.000 mm, yshift = 0.000 mm },
		boxed title style = { boxrule = 0.000 pt, sharp corners = all },
		colbacktitle = RedPurple!90!black,
		coltitle = white,
		fonttitle = \bfseries,
		varwidth boxed title,
		colback = RedPurple!03,
		colframe = RedPurple,
		sharp corners = all,
		toprule = 0.000 mm,
		bottomrule = 0.500 mm,
		leftrule = 0.500 mm,
		rightrule = 0.000 mm,
		title = {Corollary\if\relax\detokenize{#1}\relax\else~#1\fi},
	}
\newcounter{MyDefinition}[section]
\renewcommand{\theMyDefinition}{\thesection.\arabic{MyDefinition}}
\newtcolorbox{definition} [ 1 ] [ ]
	{
		breakable,
		enhanced,
		enhanced jigsaw,
		skin = enhanced,
		attach boxed title to top left = { xshift = -5.000 mm, yshift = 0.000 mm },
		boxed title style = { boxrule = 0.000 pt, sharp corners = all },
		colbacktitle = SkyBlue!70,
		coltitle = SkyBlue!30!black,
		fonttitle = \bfseries,
		varwidth boxed title,
		colback = SkyBlue!15,
		colframe = SkyBlue,
		sharp corners = all,
		toprule = 0.000 mm,
		bottomrule = 0.500 mm,
		leftrule = 0.500 mm,
		rightrule = 0.000 mm,
		code = { \refstepcounter{MyDefinition} },
		title = {Definition~\theMyDefinition:\if\relax\detokenize{#1}\relax\else~#1\fi},
	}
\newtcolorbox{definition*} [ 1 ] [ ]
	{
		breakable,
		enhanced,
		enhanced jigsaw,
		skin = enhanced,
		attach boxed title to top left = { xshift = -5.000 mm, yshift = 0.000 mm },
		boxed title style = { boxrule = 0.000 pt, sharp corners = all },
		colbacktitle = SkyBlue!70,
		coltitle = SkyBlue!30!black,
		fonttitle = \bfseries,
		varwidth boxed title,
		colback = SkyBlue!15,
		colframe = SkyBlue,
		sharp corners = all,
		toprule = 0.000 mm,
		bottomrule = 0.500 mm,
		leftrule = 0.500 mm,
		rightrule = 0.000 mm,
		title = {Definition\if\relax\detokenize{#1}\relax\else~#1\fi},
	}
\newcounter{MyExample}[section]
\renewcommand{\theMyExample}{\thesection.\arabic{MyExample}}
\newtcolorbox{example} [ 1 ] [ ]
	{
		breakable,
		enhanced,
		enhanced jigsaw,
		skin = enhanced,
		attach boxed title to top left = { xshift = -5.000 mm, yshift = 0.000 mm },
		boxed title style = { boxrule = 0.000 pt, sharp corners = all },
		colbacktitle = WordAquaAccent1Darker25,
		coltitle = white,
		fonttitle = \bfseries,
		varwidth boxed title,
		colback = WordAquaAccent1Lighter80!25,
		colframe = WordAquaAccent1Darker25,
		sharp corners = all,
		toprule = 0.000 mm,
		bottomrule = 0.500 mm,
		leftrule = 0.500 mm,
		rightrule = 0.000 mm,
		code = { \refstepcounter{MyExample} },
		title = {Example~\theMyExample:\if\relax\detokenize{#1}\relax\else~#1\fi},
	}
\newtcolorbox{example*} [ 1 ] [ ]
	{
		breakable,
		enhanced,
		enhanced jigsaw,
		skin = enhanced,
		attach boxed title to top left = { xshift = -5.000 mm, yshift = 0.000 mm },
		boxed title style = { boxrule = 0.000 pt, sharp corners = all },
		colbacktitle = WordAquaAccent1Darker25,
		coltitle = white,
		fonttitle = \bfseries,
		varwidth boxed title,
		colback = WordAquaAccent1Lighter80!25,
		colframe = WordAquaAccent1Darker25,
		sharp corners = all,
		toprule = 0.000 mm,
		bottomrule = 0.500 mm,
		leftrule = 0.500 mm,
		rightrule = 0.000 mm,
		title = {Example\if\relax\detokenize{#1}\relax\else~#1\fi},
	}
\newcounter{MyLemma}[section]
\renewcommand{\theMyLemma}{\thesection.\arabic{MyLemma}}
\newtcolorbox{lemma} [ 1 ] [ ]
	{
		breakable,
		enhanced,
		enhanced jigsaw,
		skin = enhanced,
		attach boxed title to top left = { xshift = -5.000 mm, yshift = 0.000 mm },
		boxed title style = { boxrule = 0.000 pt, sharp corners = all },
		colbacktitle = PaleVioletRed3!50,
		coltitle = black,
		fonttitle = \bfseries,
		varwidth boxed title,
		colback = WordPinkAccent1Lighter80!12,
		colframe = WordPinkAccent1Darker50,
		sharp corners = all,
		toprule = 0.000 mm,
		bottomrule = 0.500 mm,
		leftrule = 0.500 mm,
		rightrule = 0.000 mm,
		code = { \refstepcounter{MyLemma} },
		title = {Lemma~\the\theMyLemma:\if\relax\detokenize{#1}\relax\else~#1\fi},
	}
\newtcolorbox{lemma*} [ 1 ] [ ]
	{
		breakable,
		enhanced,
		enhanced jigsaw,
		skin = enhanced,
		attach boxed title to top left = { xshift = -5.000 mm, yshift = 0.000 mm },
		boxed title style = { boxrule = 0.000 pt, sharp corners = all },
		colbacktitle = PaleVioletRed3!50,
		coltitle = black,
		fonttitle = \bfseries,
		varwidth boxed title,
		colback = WordPinkAccent1Lighter80!12,
		colframe = WordPinkAccent1Darker50,
		sharp corners = all,
		toprule = 0.000 mm,
		bottomrule = 0.500 mm,
		leftrule = 0.500 mm,
		rightrule = 0.000 mm,
		title = {Lemma\if\relax\detokenize{#1}\relax\else~#1\fi},
	}
\newcounter{MyProposition}[section]
\renewcommand{\theMyProposition}{\thesection.\arabic{MyProposition}}
\newtcolorbox{proposition} [ 1 ] [ ]
	{
		breakable,
		enhanced,
		enhanced jigsaw,
		skin = enhanced,
		attach boxed title to top left = { xshift = -5.000 mm, yshift = 0.000 mm },
		boxed title style = { boxrule = 0.000 pt, sharp corners = all },
		colbacktitle = cyan7!50,
		coltitle = black,
		fonttitle = \bfseries,
		varwidth boxed title,
		colback = cyan9!25,
		colframe = cyan5,
		sharp corners = all,
		toprule = 0.000 mm,
		bottomrule = 0.500 mm,
		leftrule = 0.500 mm,
		rightrule = 0.000 mm,
		code = { \refstepcounter{MyProposition} },
		title = {Proposition~\theMyProposition:\if\relax\detokenize{#1}\relax\else~#1\fi},
	}
\newtcolorbox{proposition*} [ 1 ] [ ]
	{
		breakable,
		enhanced,
		enhanced jigsaw,
		skin = enhanced,
		attach boxed title to top left = { xshift = -5.000 mm, yshift = 0.000 mm },
		boxed title style = { boxrule = 0.000 pt, sharp corners = all },
		colbacktitle = cyan7!50,
		coltitle = black,
		fonttitle = \bfseries,
		varwidth boxed title,
		colback = cyan9!25,
		colframe = cyan5,
		sharp corners = all,
		toprule = 0.000 mm,
		bottomrule = 0.500 mm,
		leftrule = 0.500 mm,
		rightrule = 0.000 mm,
		title = {Proposition\if\relax\detokenize{#1}\relax\else~#1\fi},
	}
\newcounter{MyTheorem}[section]
\renewcommand{\theMyTheorem}{\thesection.\arabic{MyTheorem}}
\newtcolorbox{theorem} [ 1 ] [ ]
	{
		breakable,
		enhanced,
		enhanced jigsaw,
		skin = enhanced,
		attach boxed title to top left = { xshift = -5.000 mm, yshift = 0.000 mm },
		boxed title style = { boxrule = 0.000 pt, sharp corners = all },
		colbacktitle = WordAquaAccent2Darker25,
		coltitle = white,
		fonttitle = \bfseries,
		varwidth boxed title,
		colback = WordAquaAccent2Lighter80,
		colframe = WordAquaAccent2Darker25,
		sharp corners = all,
		toprule = 0.000 mm,
		bottomrule = 0.500 mm,
		leftrule = 0.500 mm,
		rightrule = 0.000 mm,
		code = { \refstepcounter{MyTheorem} },
		title = {Theorem~\theMyTheorem:\if\relax\detokenize{#1}\relax\else~#1\fi},
	}
\newtcolorbox{theorem*} [ 1 ] [ ]
	{
		breakable,
		enhanced,
		enhanced jigsaw,
		skin = enhanced,
		attach boxed title to top left = { xshift = -5.000 mm, yshift = 0.000 mm },
		boxed title style = { boxrule = 0.000 pt, sharp corners = all },
		colbacktitle = WordAquaAccent2Darker25,
		coltitle = white,
		fonttitle = \bfseries,
		varwidth boxed title,
		colback = WordAquaAccent2Lighter80,
		colframe = WordAquaAccent2Darker25,
		sharp corners = all,
		toprule = 0.000 mm,
		bottomrule = 0.500 mm,
		leftrule = 0.500 mm,
		rightrule = 0.000 mm,
		title = {Theorem\if\relax\detokenize{#1}\relax\else~#1\fi},
	}
\newcounter{mathseed}
\pgfmathsetseed{\arabic{mathseed}}
\pgfdeclarelayer{background}
\pgfsetlayers{background, main}
\pgfdeclaredecoration{irregular fractal line}{init}
{
	\state{init}[width=\pgfdecoratedinputsegmentremainingdistance]
	{
		\pgfpathlineto{\pgfpoint{random*\pgfdecoratedinputsegmentremainingdistance}{(random*\pgfdecorationsegmentamplitude-0.02)*\pgfdecoratedinputsegmentremainingdistance}}
		\pgfpathlineto{\pgfpoint{\pgfdecoratedinputsegmentremainingdistance}{0pt}}
	}
}
\def\tornpaper#1{%
	\ifthenelse{\isodd{\value{mathseed}}}
	{%
		\tikz
		{
			\node[inner sep = 1em] (A) {#1};		
			\begin{pgfonlayer}{background}			
				\fill[paper]						
				\pgfextra{\pgfmathsetseed{\arabic{mathseed}}\addtocounter{mathseed}{1}}%
				{decorate[irregular cloudy border]{decorate{decorate{decorate{decorate[ragged border]{
										(A.north west) -- (A.north east)
				}}}}}}
				-- (A.south east)
				\pgfextra{\pgfmathsetseed{\arabic{mathseed}}}%
				{decorate[irregular spiky border]{decorate{decorate{decorate{decorate[ragged border]{
										-- (A.south west)
				}}}}}}
				-- (A.north west);
			\end{pgfonlayer}
		}
	}
	{%
		\tikz{
			\node[inner sep=1em] (A) {#1};  
			\begin{pgfonlayer}{background}  
				\fill[paper] 
				\pgfextra{\pgfmathsetseed{\arabic{mathseed}}\addtocounter{mathseed}{1}}%
				{decorate[irregular spiky border]{decorate{decorate{decorate{decorate[ragged border]{
										(A.north east) -- (A.north west)
				}}}}}}
				-- (A.south west)
				\pgfextra{\pgfmathsetseed{\arabic{mathseed}}}%
				{decorate[irregular cloudy border]{decorate{decorate{decorate{decorate[ragged border]{
										-- (A.south east)
				}}}}}}
				-- (A.north east);
		\end{pgfonlayer}}
	}
}
\usepackage [ scale = 2 ] {ccicons}
\usepackage{changepage}
\usepackage{standalone}                                                                            %
\title
	{
		Evaluation of Variational Quantum Classifiers (VQC) for Cyberattack Detection in the NISQ Era
	}

\usepackage{orcidlink}
\newcommand{\orcidicon}[1]{\href{https://orcid.org/#1}{\includegraphics[height=\fontcharht\font`\B]{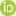}}}

\author
	{
		Angelos Thomos$^1$\orcidicon{0009-0003-1340-8515}
		and
		Theodore Andronikos$^1$\orcidicon{0000-0002-3741-1271}
		\\[10 pt]
		$^1$
		Department of Informatics, Ionian University, \\
		7 Tsirigoti Square, 49100 Corfu, Greece; \\
		\{p19thom, andronikos\}@ionio.gr
	}
\begin{document}

\maketitle

\begin{abstract}
	This paper investigates the effectiveness and structural limits of Variational Quantum Classifiers (VQC) for detecting network anomalies in the era of Noisy Intermediate-Scale Quantum (NISQ) systems. Using the official 20\% research subset of the NSL-KDD dataset, a 4-qubit classifier featuring a 24-parameter trainable ansatz was developed, utilizing amplitude encoding to embed 16 principal components. On a held-out partition of this subset, the model achieved a consistent binary-classification accuracy of 88\%. A comparative evaluation with two fundamentally different optimizers, COBYLA and SPSA, found that the observed performance plateau is not attributable to convergence to local minima, a result we interpret as consistent with an encoding-related expressibility limit rather than an optimization artifact. An architectural parity comparison with a classical neural network (a Tiny MLP with four nodes, achieving 97\% accuracy) highlighted the expressiveness gap associated with data overcompression into restricted quantum states. As a complementary diagnostic, the model was trained on a class-balanced set spanning the 22 raw NSL-KDD attack categories and evaluated in-sample (i.e., on its own training data) as a probe of representational capacity: under this configuration the VQC reached only 9\% accuracy and exhibited a degenerate mode collapse onto a small subset of classes. We interpret this behavior as consistent with the loss of linear separability induced by excessive compression in the quantum probability space, while explicitly noting that our experiments do not isolate the encoding from the ansatz depth, optimizer budget, and measurement-decoding scheme (see Limitations). Motivated by these observations and by Cover's theorem, we outline---but do not yet evaluate---an alternative paradigm: a 16-qubit VQC with angle encoding that expands the Hilbert-space representation rather than relying on aggressive classical dimensionality reduction.
	\\[12 pt]
\end{abstract}
\section{Introduction} \label{sec: Introduction}

The growing sophistication of network-based attacks has exposed a structural limitation of traditional Intrusion Detection Systems (IDS): signature-based mechanisms are inherently reactive, relying on prior knowledge of attack patterns to detect them. As adversaries adapt and novel attack vectors proliferate, the field has shifted toward machine-learning-based anomaly detection—approaches where models learn statistical baselines of normal traffic and flag deviations, as proven by Ali et al. \cite{Ali2025}, regardless of whether those deviations have been previously catalogued. The NSL-KDD dataset \cite{Hassan2020}, developed to correct the duplicate-record bias and evaluation artifacts of the earlier KDD Cup '99 benchmark, has become the canonical testbed for this research direction, offering a stratified multi-class distribution across 11 traffic categories that serves as a rigorous stress test for classifier robustness under class imbalance \cite{Buda2018, Luque2019}.

The emergence of Quantum Machine Learning (QML) has introduced a theoretically compelling alternative computational paradigm for classification tasks. By embedding classical data into exponentially high-dimensional Hilbert spaces via quantum feature maps, Variational Quantum Classifiers (VQCs)—parameterized quantum circuits trained end-to-end using classical optimization—can in principle achieve non-linear class separability that would require exponentially greater resources to reproduce classically \cite{Cerezo2021, Havlicek2019}. This theoretical promise has motivated growing interest in applying VQCs to practical machine-learning benchmarks \cite{BonetMonroig2023}, including hybrid architectures utilizing tensor networks \cite{Chen2020}.

The translation of theoretical properties into empirical performance is severely constrained by the current state of quantum hardware. The Noisy Intermediate-Scale Quantum (NISQ) era \cite{Barligea2023, Lamichhane2025} is defined by limited qubit counts, decoherence, and gate errors. These hardware limitations compound into severe optimization and encoding bottlenecks that must be carefully managed to extract any practical utility from parameterized circuits \cite{Kandala2017}.

Let us note that there exists a category of quantum algorithms that can classify functions exhibiting specific properties with probability $1.0$. In a certain sense, the Deutsch-Jozsa algorithm \cite{Deutsch1992} pioneered this field, followed by extensions such as multidimensional variants \cite{Cleve1998}, generalizations of balanced functions \cite{Chi2001, Holmes2003}, and further extension \cite{Ballhysa2004, Qiu2018, OssorioCastillo2023}. Recent works such as \cite{Andronikos2025a} and \cite{Andronikos2025b} targeted imbalanced functions, proposing the Boolean Quantum Classifiers for classifying functions with specific behavioral patterns. For a very recent probabilistic analysis of the classification outcomes with respect to the Hamming distance see also \cite{Andronikos2026}. Algorithms of the latter category typically have the form of a quantum game featuring Alice and Bob, using the engaging nature of games to clarify complex concepts. Quantum games, popularized since 1999 \cite{Meyer1999, Eisert1999}, often outperform classical strategies \cite{Andronikos2018,Andronikos2021,Andronikos2022a}, as exemplified by the Prisoners’ Dilemma \cite{Eisert1999} and other abstract games \cite{Koh2024}. Beyond entertainment, quantum games have addressed serious problems like cryptographic protocols \cite{Bennett1984, Ampatzis2021, Ampatzis2022,  Ampatzis2023, Andronikos2023, Andronikos2023a, Andronikos2023b, Karananou2024, Andronikos2024, Andronikos2024a, Andronikos2024b, Andronikos2025, Andronikos2025c}, and quantum classification of Boolean functions \cite{Andronikos2025a, Andronikos2025b}. Furthermore, many classical systems can be transformed into quantum versions, including political frameworks, as demonstrated in recent studies \cite{Andronikos2022}. Games make be envisioned in unconventional situations, such as those featuring biological systems, which have actually attracted a lot of attention during the last decade \cite{Theocharopoulou2019, Kastampolidou2020a, Kostadimas2021, Polenakis2026}. It may even be the case to discover biosystems that employ biostrategies that perform better than conventional strategies—even in the well-known Prisoners' Dilemma game (see for example \cite{Kastampolidou2020, Kastampolidou2021, Papalitsas2021, Kastampolidou2023, Adam2023}).

Applying QML to network intrusion detection introduces a critical challenge specific to this encoding step. High-dimensional traffic data must be compressed before it can be loaded into a restricted qubit register. Amplitude Encoding—which maps $2^n$ real-valued features into the amplitude vector of an $n$-qubit state—achieves high representational density, but at the cost of lossy compression \cite{Barligea2023, Ranga2024}. Whether this compression constitutes a recoverable preprocessing constraint or an irreducible structural bottleneck is the empirical question this paper exists to answer. The central research question investigated here is therefore:

\begin{quote}
	\textcolor{darkgray}{\textit{Can a Variational Quantum Classifier with Amplitude Encoding on 4 qubits constitute an effective and competitive solution for network anomaly detection on the NSL-KDD dataset, or does its architecture impose a structural limit on its performance?}}
\end{quote}

This question is deliberately constructed so that both outcomes are scientifically informative. A positive result would establish a methodological template for quantum-assisted IDS. A negative result carries equal scientific value: by identifying precisely where and why the architecture fails, we provide the empirical constraints required to move beyond premature claims of advantage. Our experiments yield a definitive, quantitative answer to this question, mapping a specific failure mode with constructive implications.

\subsection{Related work} \label{sec: Related Work}

The integration of QML into cybersecurity represents one of the most promising frontiers in modern network defense. A primary focus of foundational research has been navigating the structural complexities of standardized benchmarks like the NSL-KDD dataset \cite{Tavallaee2009, Hassan2020}. While classical machine learning has established highly effective mechanisms for handling the imbalanced distributions inherent to network traffic \cite{Buda2018, Luque2019}, with recent deterministic studies solidifying the robustness of interpretable classical baselines like Logistic Regression on the NSL-KDD dataset \cite{ArcosArgudo2025}, researchers have successfully begun translating these challenges into quantum states for diverse cybersecurity applications \cite{Havlicek2019, Alghaythi2022, Umer2024, Sarvade2025, Wichert2024}.

A major success of recent literature has been the engineering ingenuity used to deploy VQCs within strict physical NISQ limitations \cite{Xu2024, Li2024}. Because current quantum hardware cannot natively process the 41 features of NSL-KDD, researchers have proactively addressed feature loss through diverse strategies, such as hybrid QCNNs \cite{Xu2024} and successful dimension reduction via successive Schmidt decompositions \cite{Daskin2023}. These pragmatic adaptations have allowed the field to test hybrid quantum-classical models on real-world data long before fault-tolerant computers become available.

As these architectures have matured, empirical studies have carefully documented their natural performance boundaries \cite{Kwon2025}. Recent comparative analyses in cybersecurity further corroborate this limitation, demonstrating that while quantum frameworks show theoretical promise, parameterized quantum architectures (such as QCNNs) often struggle to match the performance of classical neural networks due to the severe hardware constraints of the NISQ era \cite{Eze2025}. In exploring these boundaries, the community has investigated various factors, ranging from the expressibility limits of parameterized circuits \cite{Kandala2017} to the severe optimization challenges posed by Barren Plateaus—regions of near-zero gradient where traditional algorithms fail \cite{Kulshrestha2022}. This has led to a rich ongoing discussion regarding optimizer selection, with researchers actively comparing gradient-free methods like COBYLA \cite{Powell1994} against stochastic approximations like SPSA \cite{Spall1998} to maximize VQA performance \cite{BonetMonroig2023}.

Building directly upon this substantial body of work, our research takes the next logical step. The community has demonstrated how to compress and encode network data, and has mapped the resulting performance characteristics. What remains to be explored is a precise mathematical isolation of the factors contributing to these boundaries. By benchmarking deterministic optimization (COBYLA) against stochastic perturbation (SPSA) on a strictly controlled architecture, this paper isolates the optimization variable to clearly understand how data compression interacts with quantum encoding.

\textbf{Contribution}. This research establishes an empirical baseline for Quantum Machine Learning (QML) in network intrusion detection by synthesizing theoretical quantum methodologies with the physical realities of Noisy Intermediate-Scale Quantum (NISQ) devices. By systematically stress-testing this pipeline, the present work isolates the specific structural vulnerabilities of applying linear compression and low-expressibility encodings to complex network traffic. Our contributions can be summarized as follows.

\begin{itemize}
	\item	
	\textbf{Experimental Results.} We provide extensive experimental results consistent with the theoretical finding by Kadi et al. \cite{Kadi2025} that Amplitude Embedding suffers from severely restricted expressibility. By compressing 16 features into the state space of just 4 qubits, we observe a representational bottleneck: when trained on a class-balanced set spanning the 22 raw NSL-KDD attack categories and evaluated in-sample, the model undergoes a degenerate mode collapse, with accuracy plateauing at 9\%. Because the model fails to fit even the training data it was optimized on, this behaviour is suggestive of a capacity limitation rather than a pure generalization failure. We are explicit, however, that this single configuration does not by itself isolate the encoding as the sole cause: the ansatz depth, the optimizer budget, and the measurement-to-label decoding scheme are plausible co-contributors, and we treat the encoding as the most likely---but not conclusively isolated---driver (see subsection \ref{subsec: Limitations, Threats To Validity & Future Directions}). We therefore make no claim that the 9\% figure constitutes a formal mathematical bound; we report it as a reproducible empirical observation consistent with a practical ceiling imposed by the loss of class-separating information during Amplitude Encoding. As a constructive response, we propose moving beyond Amplitude Embedding for this class of problems and outline an alternative architecture: an Angle-Entangled VQC employing 16 qubits.
	\item	
	\textbf{Evaluation of Principal Component Analysis.} Building upon the foundational recommendation by Lamichhane \& Rawat \cite{Lamichhane2025} to utilize dimensionality reduction as a necessary facilitation for NISQ devices, we empirically evaluated the application of classical Principal Component Analysis (PCA). We also strongly support the premise that dimensionality reduction is essential in the current hardware era; however, the application of PCA (compressing the NSL-KDD dataset from 41 to 16 dimensions) revealed a specific operational boundary. We conjecture that compression methods run the risk of eliminating variance components critical for minority attack classes, thereby degrading linear separability. Without doubt, the need for preprocessing is imperative, but it seems that techniques specific to the quantum classification task at hand must be developed, such as quantum-native, reduction strategies.
	\item	
	\textbf{Divergence between Theory \& Practice.} Inspired by the theoretical performance ceilings and the promise of quantum advantage, as articulated in seminal works such as Iovane \cite{Iovane2025}, this paper investigates how closely these theoretical ideals can be approximated under current, noise-emulated NISQ constraints. To establish a rigorous empirical baseline, we evaluated the quantum model against a classical ``Tiny MLP'' with a four-node hidden layer---a strict architectural analog to the 4-qubit space. Under identical input compression, the classical model achieved 97\% accuracy, exceeding the VQC's 88\%. This divergence between theoretical potential and current empirical performance does not diminish the profound promise of quantum models. Instead, it constructively maps a specific methodological pathway that currently falls short, revealing that realizing the theoretical optimism of QML requires moving beyond aggressive classical dimensionality reduction and low-expressibility encodings.
	\item	
	\textbf{Angle Encoding as the Most Viable Path Forward.} Acknowledging the inherent bottlenecks of compressed quantum feature maps fundamentally motivates a paradigm shift toward scalable, high-fidelity Quantum Machine Learning architectures. To decisively circumvent the ``Quantum Bottleneck,'' an uncompressed encoding strategy is not merely beneficial—it is essential. We therefore advocate a direct ``One Feature per Qubit'' framework based on angle encoding, in which each classical feature is embedded. Such a design would preserve the full granularity of the input data while simultaneously exploiting Cover's Theorem on pattern separability. By projecting the classical feature vector into a quantum Hilbert space whose dimensionality expands exponentially with the number of qubits, we ensure that the embedded data resides in a regime where Cover's theorem guarantees that the probability of linear separability approaches unity. In practice, this exponential lift renders even highly entangled, non-linear attack vectors overwhelmingly likely to be linearly separable.
\end{itemize}
\subsection*{Organization} \label{subsec: Organization}

This article is organized as follows: Section \ref{sec: Introduction} introduces the topic, while Section \ref{sec: Related Work} discusses prior work. Section \ref{sec: Implementation Environment & Reproducibility} details the implementation setup. Section \ref{sec: NSL-KDD Dataset Analysis} analyzes the dataset, and Section \ref{sec: Data Preprocessing & Feature Engineering} outlines the preprocessing and feature engineering steps. Section \ref{sec: Model Architecture} maps the quantum circuit design, and Section \ref{sec: Optimization Landscape & Methodological Safeguards} explains the optimization strategy and methodological safeguards. Sections \ref{sec: Binary Classification Results} and \ref{sec: Multiclass Classification: Mapping The Boundaries Of Amplitude Encoding} present the binary and multi-class experimental results, respectively. Finally, Section \ref{sec: Discussion & Conclusions} provides the concluding remarks and the future roadmap.

\section{Implementation environment \& reproducibility} \label{sec: Implementation Environment & Reproducibility}

In view of the fact that the Qiskit ecosystem evolves and restructures so quickly, strict reproducibility was prioritized. Development and simulation took place within an isolated Python Virtual Environment. The deliberate choice to utilize a statevector simulator (\texttt{Qiskit-Aer}) rather than a physical Quantum Processing Unit (QPU) was made to strictly isolate the structural expressiveness of the quantum encoding. Executing on real hardware would introduce decoherence and gate errors, making it impossible to definitively separate the intrinsic mathematical limits of amplitude encoding from hardware-induced noise. The host machine utilized an AMD Ryzen 5 3600 processor, 16GB of RAM, and an Intel ARC B570 10GB GPU running Arch Linux.

The following IBM ecosystem components were utilized:

\begin{itemize}
	\item	
	Qiskit 2.3.0
	\item	
	Qiskit-Aer 0.17.2 (simulator)
	\item	
	Qiskit-Machine-Learning 0.9.0
	\item	
	Qiskit-Algorithms 0.4.0
\end{itemize}

For data preprocessing, analysis, and visualization, the following libraries were utilized:

\begin{itemize}
	\item	
	Pandas (3.0.1)
	\item	
	NumPy (2.4.3)
	\item	
	Scikit-Learn (1.8.0)
	\item	
	SciPy (1.17.1)
	\item	
	Matplotlib (3.10.8)
	\item	
	Seaborn (0.13.2)
\end{itemize}

To guarantee complete transparency and allow for the exact reproduction of the reported metrics, the entire source code—including the dataset preprocessing pipelines, custom optimization callbacks, and the complete quantum circuit implementations—has been made publicly available in the project GitHub repository: \url{https://github.com/Angeloth1/Quantum-IDS-NSLKDD}. The repository's \texttt{README} documents which branch reproduces which table and figure in this paper, since several methodological variants (label granularity, evaluation protocol) were developed on separate branches during the iterative research process.

\section{NSL-KDD Dataset Analysis} \label{sec: NSL-KDD Dataset Analysis}

The NSL-KDD dataset~\cite{Hassan2020} was selected for this study as a refined, rigorous benchmark for intrusion detection, carefully evaluated against the specific hardware constraints of the Noisy Intermediate-Scale Quantum (NISQ) era.

\subsection{Dataset architecture \& sampling strategy} \label{subsec: Dataset Architecture & Sampling Strategy}

NSL-KDD records complete network flows, where each record is defined by 41 features logically grouped into three categories: Basic Features (e.g., duration, protocol, byte volume), Content Features (payload data for detecting suspicious behavior), and Traffic Features (time and host statistics critical for identifying overload attacks). The recorded attacks are classified into four main families: Denial of Service (DoS), Probe (Scanning), Remote to Local (R2L), and User to Root (U2R) \cite{Tavallaee2009, Hassan2020}. This structured taxonomy allows the quantum model to be evaluated across diverse threat scenarios.

As visually demonstrated by the t-SNE projection of the 20\% research subset in Figure \ref{fig: fig1}, the dataset exhibits highly complex, non-linear overlaps between different traffic classes. Specifically, while DoS and Normal traffic form massive, relatively distinct manifolds, minority attacks such as U2R and R2L are exceptionally sparse and heavily entangled within the normal traffic distribution. The difficulty of classifying these imbalanced, deeply entangled structures using traditional models without aggressive manipulation has been robustly documented in recent deterministic comparisons on the NSL-KDD \cite{ArcosArgudo2025}. Consequently, this visual and statistical evidence strongly justifies the exploration of exponentially high-dimensional quantum feature spaces (Hilbert spaces) to theoretically unlock the non-linear separability required to isolate these microscopic attack vectors \cite{Havlicek2019}.

\begin{tcolorbox}
	[
	enhanced,
	breakable,
	grow to left by = 1.500 cm,
	grow to right by = 0.000 cm,
	colback = white,
	enhanced jigsaw,			
	frame hidden,
	sharp corners,
	]
	\begin{figure}[H]
		\centering
		\includegraphics [ scale = 0.500, trim = {0.000cm 0.000cm 0.000cm 0.000cm}, clip ] {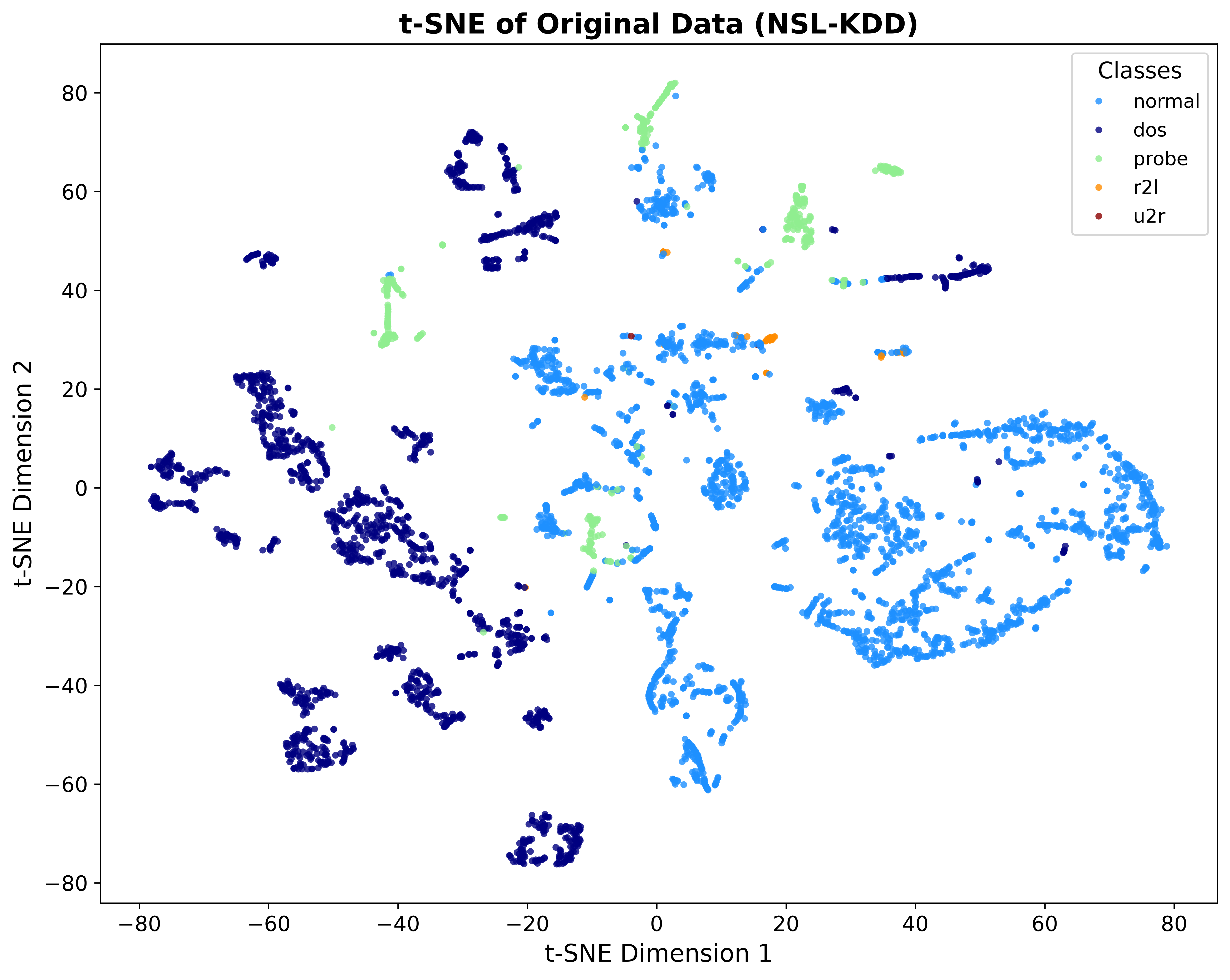}
		\caption{Visualizing the complex dataset distribution using a t-SNE projection on the official 20\% NSL-KDD research subset.}
		\label{fig: fig1}
	\end{figure}
\end{tcolorbox}

Due to the extreme computational complexity of simulating Parameterized Quantum Circuits (PQCs) via classical statevector simulators, training on the entire 125,973-record dataset was deemed unfeasible. Consequently, the official 20\% research subset comprising 25,192 records (\texttt{KDDTrain+\_20Percent}) was adopted. This subset was further partitioned using an 80/20 train-test split (20,155 records for training and 5,037 for testing), carefully preserving the underlying attack distributions. We emphasize that all results in this study are obtained on held-out partitions of this 20\% subset; the separate official evaluation file (\texttt{KDDTest+}), which introduces attack types unseen during training, is deliberately outside the scope of this work due to the same simulation-cost constraints. Accordingly, our accuracy figures should be read as characterizing behaviour \emph{within} the 20\% subset rather than as estimates of generalization to the full NSL-KDD benchmark; evaluation on \texttt{KDDTest+} is identified as a priority for future work.

\subsection{Dataset selection \& the quantum bottleneck} \label{subsec: Dataset Selection & the Quantum Bottleneck}

The selection of NSL-KDD over other established benchmarks is intrinsically tied to the computational realities of current quantum hardware. Compared to its predecessor, KDD Cup '99, NSL-KDD eliminates the massive redundancy (roughly 75–78\% duplicate records) that previously biased classifiers toward frequent attack types, thereby offering a more statistically rigorous stress test \cite{Tavallaee2009}. Moreover, recent comparative literature has established strong, interpretable classical baselines (such as Logistic Regression) on the NSL-KDD, allowing for highly deterministic evaluations of new hybrid models \cite{ArcosArgudo2025}.

However, the primary defining factor for dataset selection remains the ``Quantum Bottleneck.'' While modern datasets such as UNSW-NB15 or CICIDS2017 incorporate highly contemporary botnet topologies, their massive feature spaces (often exceeding 80 variables) and multi-million record volumes render them fundamentally incompatible with NISQ-era resources. Recent evaluations in malicious URL detection clearly corroborate that while quantum computation offers theoretical advantages via high-dimensional feature spaces, practical implementations of Quantum Neural Networks struggle to scale effectively against classical CNNs precisely due to current hardware limitations \cite{Eze2025}. To process 80+ features, a VQC would either require a prohibitive number of qubits or mandate extreme classical dimensionality reduction, which irreparably destroys the data's variance. With its native 41 features, NSL-KDD strikes an optimal balance: it is complex enough to challenge the classifier, yet allows for a manageable compression phase (from 41 down to 16 features) that prevents the model from completely collapsing prior to quantum encoding.

\section{Data preprocessing \& feature engineering} \label{sec: Data Preprocessing & Feature Engineering}

Data preprocessing is a critical transitional stage because it transforms raw, highly nonlinear network records into a compact, strictly scaled mathematical format, ensuring compatibility with the geometric constraints of a Variational Quantum Classifier (VQC).

\subsection{Data segmentation \& label standardization} \label{subsec: Data Segmentation & Label Standardization}

Since the primary operational goal of a foundational IDS is the immediate and reliable detection of threats, the 22 distinct attack categories within the NSL-KDD dataset were aggregated into a single malicious class (Class 1), with normal traffic designated as Class 0. This label binarization allows the VQC to focus exclusively on learning the optimal decision boundary between safe and malicious states, eliminating the statistical noise associated with highly imbalanced subcategories. For data segmentation, a simple target-label stratification was deemed insufficient. Instead, a dual-stratification strategy was implemented, combining the target label with the payload size variable \texttt{src\_bytes}. Packet size constitutes a critical physical signature: DoS attacks are structurally characterized by the rapid transmission of microscopic packets, whereas U2R vectors necessitate the transfer of large, anomalous files. To handle its massive variance, the continuous \texttt{src\_bytes} variable was mathematically converted into a categorical feature via equal-frequency quantiles (low, medium, high).

By splitting the dataset along this composite axis, the risk of distributional shift is aggressively mitigated. During this process, exactly 8 attack combinations were identified as absolute outliers (appearing only once in the entire dataset). To prevent data loss and ensure the model was exposed to these rare signatures, these specific outliers were manually isolated and explicitly injected into the Training Set. The remaining majority of the data was then rigorously stratified using an 80/20 train-test split, ensuring the classifier would not memorize incorrect statistical correlations between packet volume and threat probability.

\subsection{Quantum-compliant encoding \& dimensionality reduction} \label{subsec: Quantum-Compliant Encoding & Dimensionality Reduction}

Converting complex categorical attributes into numerical vectors introduces a well-documented bottleneck for quantum algorithms. We did not adopt the traditional classical approach One-Hot Encoding (OHE), since we estimated that applying OHE to the NSL-KDD dataset would generate over 100 new, highly sparse columns. In a Noisy Intermediate-Scale Quantum (NISQ) environment, embedding such extreme sparsity would mandate excessively deep, parameter-heavy circuits, thereby multiplying gate errors and decoherence rates.

Instead, Target Encoding was adopted, replacing each categorical value (such as protocol type and the newly created source bytes category) with the historical posterior probability of an attack occurring \cite{Nadeem2026}. This methodological shift offers a dual advantage: it directly embeds semantic risk value into the feature, and it provides strict ``Quantum Economy,'' condensing the categorical dimensions seamlessly without generating sparse matrices. 

To strictly prevent data leakage—a critical vulnerability in predictive cybersecurity modeling—all preprocessing transformations were subjected to a rigid, deterministic pipeline \cite{ArcosArgudo2025}. The Target Encoder (incorporating automatic smoothing to prevent overfitting on rare classes) and the Standard Scaler (centering data to zero mean and unit variance) were exclusively fitted to the Training Set. The Test Set was subsequently transformed using only these locked parameters.

In the final dimensionality reduction stage, Principal Component Analysis (PCA \cite{Jolliffe2002}) was applied to extract exactly 16 principal components. This specific threshold was deliberately chosen to fully saturate a 4-qubit Hilbert space via Amplitude Encoding ($2^4 = 16$), without introducing artifacts via zero-padding. Evaluated empirically, this PCA compression successfully preserved 86.78\% of the dataset's original explained variance. Ultimately, the synergistic pipeline of Target Encoding, rigorous standard scaling, and PCA-16 successfully reduced the quantum resource requirements while preserving the structural integrity of the cyberattack signatures.

\section{Model architecture} \label{sec: Model Architecture}

To enable rigorous benchmarking of different optimization algorithms, a detailed anatomical description of the implemented Variational Quantum Classifier (VQC) is necessary. The structural design of the circuit directly dictates the topological depth, the number of trainable parameters, and, consequently, the complexity of the loss landscape that the optimizers must navigate.

The overall quantum circuit is structured into three distinct, sequential stages: data encoding (State Preparation), the trainable parameterized block (Ansatz \cite{Cerezo2021}), and the final state collapse (Measurement).

\subsection{State preparation \& encoding constraints} \label{subsec: State Preparation & Encoding Constraints}

The initial stage of the circuit translates the classical vector of 16 scaled principal components into a 4-qubit quantum state (spanning registers $q_0$ to $q_3$). Prior to embedding, the classical features were strictly normalized using the $\ell_2$ norm ($\|x\|^2 = 1$). This mathematical prerequisite ensures that the sum of the squared amplitudes equals exactly 1, satisfying the fundamental postulate of quantum mechanics regarding probability conservation. The mathematical representation of the prepared input state is defined as:

\begin{align}
	\label{eq: Input State}
	\ket{ \psi_{in} }
	=
	\sum_{ i = 0 }^{ 15 }
	x_{i}
	\ket{ i }
	\ .
\end{align}

\begin{lstlisting}
	state_prep = StatePreparation(sample_data)
	qc = QuantumCircuit(
	num_qubits, name="Amplitude encoding"
	)
	qc.append(state_prep, range(num_qubits))
	feature_vector_circuit = raw_feature_vector(
	feature_dimension=16
	)
\end{lstlisting}

\begin{figure}[H]
	\centering
	\begin{tikzpicture}
		\begin{yquant}
			qubit {$q_{\idx}$} q[4];
			[fill=cyan!10, draw=cyan!80!black, thick, rounded corners=2pt] box {Parameterized\\Initialize} (q[0-3]);
			align q;
			[fill=cyan!10, draw=cyan!80!black, thick, rounded corners=2pt] box {$R_y(\theta_0)$} q[0];
			[fill=cyan!10, draw=cyan!80!black, thick, rounded corners=2pt] box {$R_y(\theta_1)$} q[1];
			[fill=cyan!10, draw=cyan!80!black, thick, rounded corners=2pt] box {$R_y(\theta_2)$} q[2];
			[fill=cyan!10, draw=cyan!80!black, thick, rounded corners=2pt] box {$R_y(\theta_3)$} q[3];
			align q;
			[style={red!70!black, thick}] cnot q[0] | q[1];
			align q;
			[style={red!70!black, thick}] cnot q[1] | q[2];
			align q;
			[style={red!70!black, thick}] cnot q[2] | q[3];
			align q;
			text {$\dots$} (q[0-3]);
			measure q[0-3];
		\end{yquant}
	\end{tikzpicture}
	\caption{VQC architecture showcasing the state preparation block followed by a RealAmplitudes-style ansatz.}
	\label{fig: QuantumCircuit}
\end{figure}
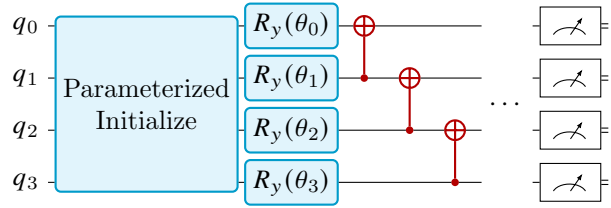

Amplitude Encoding was deliberately chosen to practically stress-test the concept of ``Quantum Economy'' \cite{Schuld2021, Stenger2025}. By leveraging the exponential nature of Hilbert space~\cite{Havlicek2019}, it successfully embeds $2^n$ classical features into merely $n$ qubits ($2^4=16$). However, due to this extreme state compression, the initialization is not a simple linear mapping. Through the \texttt{raw\_feature\_vector} class, the circuit dynamically synthesizes an extensive sequence of multi-controlled unitary gates to precisely map the classical data onto the probability amplitudes of the 16 computational basis states.

It is crucial to note that while this encoding segment remains fixed (non-trainable) during the optimization phase, recent studies confirm that amplitude state preparation requires a circuit depth that scales exponentially, $\mathcal{O}(2^n)$, in the worst-case scenario \cite{Nikam2026, Ranga2024}. This structural limitation acts as a massive computational burden and a primary source of simulated hardware noise, actively testing the robustness of the chosen optimizers.

\subsection{Trainable ansatz \& expressivity} \label{subsec: Trainable Ansatz & Expressivity}

The ``neural'' component of the VQC is structured upon the hardware-efficient \texttt{RealAmplitudes} architecture \cite{Kandala2017}. This specific ansatz was selected for its ability to generate robust entanglement while strictly confining its unitary operations to the real subspace. By eliminating complex parameters, it significantly reduces the multidimensional search space, facilitating much smoother convergence for classical optimizers like COBYLA and SPSA.

The architecture comprises repeating building blocks (\texttt{reps}), where a single repetition involves two distinct layers:

\begin{itemize}
	\item	
	\textbf{Rotation Layer:} The application of parameterized $R_{ y } ( \theta )$ gates independently to each qubit. Mathematically, the core training unit is the rotation around the Y-axis of the Bloch sphere, described by the matrix:
	\begin{align}
		\label{eq: Rotation Matrix}
		R_{ y } ( \theta )
		=
		\begin{bNiceMatrix}[ margin ] 
			\cos ( \frac { \theta } { 2 } ) & - \sin ( \frac { \theta } { 2 } ) \\
			\sin ( \frac { \theta } { 2 } ) & \phantom{-} \cos ( \frac { \theta } { 2 } ) \\
		\end{bNiceMatrix}
		\ .
	\end{align}
	\item	
	\textbf{Entanglement Layer:} A linear entanglement topology implemented sequentially via CNOT gates, ensuring logical correlation between adjacent qubits.
\end{itemize}

\begin{lstlisting}
	ansatz_amp = RealAmplitudes(
	num_qubits=4, reps=5
	)  # [...]
	vqc_amp = feature_vector_circuit.compose(
	ansatz_amp
	)
\end{lstlisting}

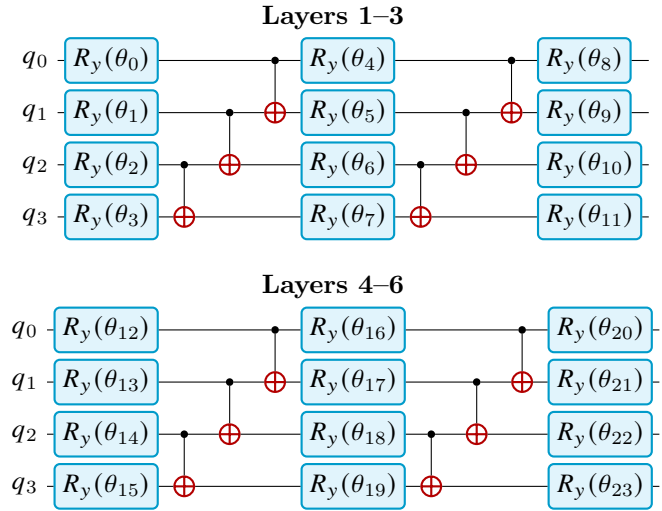
\begin{figure}[H]
	\centering
	\textbf{Layers 1--3}\\[2pt]
	\begin{tikzpicture}
		\begin{yquant}
			qubit {$q_{\idx}$} q[4];
			[fill=cyan!10, draw=cyan!80!black, thick, rounded corners=2pt] box {$R_y(\theta_0)$} q[0]; 
			[fill=cyan!10, draw=cyan!80!black, thick, rounded corners=2pt] box {$R_y(\theta_1)$} q[1]; 
			[fill=cyan!10, draw=cyan!80!black, thick, rounded corners=2pt] box {$R_y(\theta_2)$} q[2]; 
			[fill=cyan!10, draw=cyan!80!black, thick, rounded corners=2pt] box {$R_y(\theta_3)$} q[3];
			[draw=red!70!black, thick] cnot q[3] | q[2]; align q; 
			[draw=red!70!black, thick] cnot q[2] | q[1]; align q; 
			[draw=red!70!black, thick] cnot q[1] | q[0]; align q;
			[fill=cyan!10, draw=cyan!80!black, thick, rounded corners=2pt] box {$R_y(\theta_4)$} q[0]; 
			[fill=cyan!10, draw=cyan!80!black, thick, rounded corners=2pt] box {$R_y(\theta_5)$} q[1]; 
			[fill=cyan!10, draw=cyan!80!black, thick, rounded corners=2pt] box {$R_y(\theta_6)$} q[2]; 
			[fill=cyan!10, draw=cyan!80!black, thick, rounded corners=2pt] box {$R_y(\theta_7)$} q[3];
			[draw=red!70!black, thick] cnot q[3] | q[2]; align q; 
			[draw=red!70!black, thick] cnot q[2] | q[1]; align q; 
			[draw=red!70!black, thick] cnot q[1] | q[0]; align q;
			[fill=cyan!10, draw=cyan!80!black, thick, rounded corners=2pt] box {$R_y(\theta_8)$} q[0]; 
			[fill=cyan!10, draw=cyan!80!black, thick, rounded corners=2pt] box {$R_y(\theta_9)$} q[1]; 
			[fill=cyan!10, draw=cyan!80!black, thick, rounded corners=2pt] box {$R_y(\theta_{10})$} q[2]; 
			[fill=cyan!10, draw=cyan!80!black, thick, rounded corners=2pt] box {$R_y(\theta_{11})$} q[3];
		\end{yquant}
	\end{tikzpicture}
	
	\vspace{8pt}
	
	\textbf{Layers 4--6}\\[2pt]
	\begin{tikzpicture}
		\begin{yquant}
			qubit {$q_{\idx}$} q[4];
			[fill=cyan!10, draw=cyan!80!black, thick, rounded corners=2pt] box {$R_y(\theta_{12})$} q[0]; 
			[fill=cyan!10, draw=cyan!80!black, thick, rounded corners=2pt] box {$R_y(\theta_{13})$} q[1]; 
			[fill=cyan!10, draw=cyan!80!black, thick, rounded corners=2pt] box {$R_y(\theta_{14})$} q[2]; 
			[fill=cyan!10, draw=cyan!80!black, thick, rounded corners=2pt] box {$R_y(\theta_{15})$} q[3];
			[draw=red!70!black, thick] cnot q[3] | q[2]; align q; 
			[draw=red!70!black, thick] cnot q[2] | q[1]; align q; 
			[draw=red!70!black, thick] cnot q[1] | q[0]; align q;
			[fill=cyan!10, draw=cyan!80!black, thick, rounded corners=2pt] box {$R_y(\theta_{16})$} q[0]; 
			[fill=cyan!10, draw=cyan!80!black, thick, rounded corners=2pt] box {$R_y(\theta_{17})$} q[1]; 
			[fill=cyan!10, draw=cyan!80!black, thick, rounded corners=2pt] box {$R_y(\theta_{18})$} q[2]; 
			[fill=cyan!10, draw=cyan!80!black, thick, rounded corners=2pt] box {$R_y(\theta_{19})$} q[3];
			[draw=red!70!black, thick] cnot q[3] | q[2]; align q; 
			[draw=red!70!black, thick] cnot q[2] | q[1]; align q; 
			[draw=red!70!black, thick] cnot q[1] | q[0]; align q;
			[fill=cyan!10, draw=cyan!80!black, thick, rounded corners=2pt] box {$R_y(\theta_{20})$} q[0]; 
			[fill=cyan!10, draw=cyan!80!black, thick, rounded corners=2pt] box {$R_y(\theta_{21})$} q[1]; 
			[fill=cyan!10, draw=cyan!80!black, thick, rounded corners=2pt] box {$R_y(\theta_{22})$} q[2]; 
			[fill=cyan!10, draw=cyan!80!black, thick, rounded corners=2pt] box {$R_y(\theta_{23})$} q[3];
		\end{yquant}
	\end{tikzpicture}
	\caption{Detailed architecture of the RealAmplitudes ansatz (\texttt{reps=5}), highlighting the parameterized $R_y$ rotation layers and the linear entanglement topology.}
	\label{fig:ansatz_detail}
\end{figure}

For the empirical benchmarks in this study, the Ansatz was configured with \texttt{reps=5}. This results in a total of 6 rotation layers (one initial, unentangled layer followed by five repetitions) applied across the 4-qubit register. Consequently, the optimization algorithm is tasked with training a vector of exactly 24 independent parameters ($\theta_0$ to $\theta_{23}$). This specific parameter count serves as the critical complexity metric utilized later for architectural parity comparisons against classical model ~\cite{Sim2019}.

\subsection{Measurements \& label mapping} \label{subsec: Measurements & Label Mapping}

In the final computational stage, measurement operations are applied to all 4 qubits along the computational Z-basis, effectively collapsing the quantum superposition into a classical binary bitstring. Mathematically, the probability $P ( i )$ of measuring a specific ground state $\ket{ i }$ is dictated by the Born rule:

\begin{align}
	\label{eq: Measurement}
	P ( i ) = \abs{ \braket { i } { \psi_{ out } } }^{ 2 }
\end{align}

\begin{lstlisting}
	vqc_amp_model_final = VQC(...) 
\end{lstlisting}

To translate this 4-bit probability distribution into a binary classification outcome, a parity mapping function is employed (the default behavioral protocol within Qiskit's \texttt{VQC} class, widely utilized as a standard observable for evaluating bitstring outcomes in variational classifiers \cite{Sen2022}). The model evaluates the parity of the observed bitstring (the sum of its constituent bits modulo 2). If the bitstring contains an even number of $1$s (e.g., $0000$, $0011$), it is mapped to Class 0 (Normal traffic). Conversely, if the bitstring contains an odd number of $1$s (e.g., $0001$, $0111$), it is mapped to Class 1 (Malicious Attack). The class assigned the highest cumulative expectation value across all mapped bitstrings determines the ultimate prediction of the IDS.

\section{Optimization landscape \& methodological safeguards} \label{sec: Optimization Landscape & Methodological Safeguards}

With the architectural geometry of the VQC mathematically defined, the 24 parameterized weights of the Ansatz require precise calibration to map the quantum states to the correct class labels of the NSL-KDD dataset. Unlike classical neural networks, which rely almost exclusively on analytical backpropagation, extracting exact analytic gradients from a quantum circuit directly typically requires evaluating the Parameter-Shift Rule \cite{Schuld2019} or deploying advanced unitary-based gradient formulations~\cite{Urbaneja2026}. This Hybrid Quantum-Classical (HQC) process dictates that for every individual parameter, the circuit must be executed multiple times with shifted angles. Such a back-and-forth data exchange creates severe system bottlenecks and high latency \cite{Pratibha2025}, inflating the computational overhead and exposing the gradient estimations to cumulative gate errors and noise inherent to NISQ hardware \cite{DellAnna2025}.

Furthermore, heavily parameterized quantum models—especially those reliant on the exponentially deep circuits required for Amplitude Encoding—are highly susceptible to Barren Plateaus \cite{Kulshrestha2022}. These are vast, mathematically flat regions within the optimization landscape where gradients vanish exponentially, rendering traditional descent-based algorithms practically inert. To definitively ascertain whether the classification limits of the VQC stem from a fundamental lack of quantum expressivity or merely an inability to navigate this complex, noisy loss landscape, a robust comparative test strategy was implemented using two optimizers with radically opposing philosophies.

The first algorithm deployed is the Constrained Optimization BY Linear Approximations (COBYLA) \cite{Powell1994}. As a deterministic, gradient-free method \cite{BonetMonroig2023}, COBYLA circumvents the latency of parameter-shift gradient evaluations entirely. It operates by constructing a local geometric simplex around the current parameters and linearly approximating the objective function to determine the optimal descent step. While highly computationally efficient and capable of rapid convergence in noise-free simulations, its strictly localized vision makes it highly vulnerable to becoming permanently trapped in shallow local minima if the error surface exhibits sharp geometric irregularities.

To counterbalance the inherent vulnerabilities of deterministic descent, the Simultaneous Perturbation Stochastic Approximation (SPSA) \cite{Kandala2017, Spall1998} was deployed simultaneously. Developed specifically for high-noise environments, SPSA is a stochastic algorithm that approximates the global gradient by randomly perturbing all 24 model parameters simultaneously across the Hilbert space. This controlled stochastic oscillation acts as a kinetic force, enabling the optimizer to effectively bounce out of localized minima and traverse flat Barren Plateaus, making it an industry standard for NISQ-era training.

This dual-optimizer deployment is not structured as a comparative benchmark of speed, but rather as the foundational methodological safeguard of this study. In any Parameterized Quantum Circuit, sub-optimal classification accuracy arises from two mutually exclusive vectors: either an optimization failure (the algorithm is trapped and fails to discover the mathematically optimal weights) or a failure of expressivity (the geometric structure of the encoding lacks the necessary Hilbert-space complexity to separate the data points, regardless of the weights). By benchmarking the locally bounded COBYLA against the globally perturbing SPSA, the training variable is effectively isolated. If the stochastic resilience of SPSA yields exactly the same empirical performance ceiling as the deterministic COBYLA, it successfully demonstrates that the limitation is not an optimization artifact, but an intrinsic, insurmountable boundary imposed by the chosen data compression and encoding geometry.

\section{Binary classification results} \label{sec: Binary Classification Results}

To empirically evaluate the foundational capability of the Quantum Model (VQC) to distinguish between normal and malicious network traffic, a binary classification paradigm was established. The encoding of the 16 Principal Components (PCA-16) was implemented utilizing the Amplitude Encoding technique (via Qiskit’s \texttt{raw\_feature\_vector}). This specific embedding strategy successfully condensed the entire feature vector into a highly restrictive 4-qubit register, leveraging the exponential volume of the quantum Hilbert space ($2^4 = 16$).

Complementing this compact encoding, the trainable parameterized block was constructed using the \texttt{RealAmplitudes} architecture. To maximize expressiveness and effectively map the complex, non-linear correlations of the NSL-KDD dataset without overwhelming the simulation, the circuit depth was configured with \texttt{reps=5}. This structural choice generated a circuit with robust entanglement topologies \cite{Deng2017}, yielding a precise total of 24 trainable parameters responsible for delineating the decision boundaries.

\subsection{VQC training configuration \& empirical performance} \label{subsec: VQC Training Configuration & Empirical Performance}

For exact reproducibility, Table \ref{tab: config} consolidates the full experimental configuration used to produce the binary results reported below; the multiclass experiment of Section \ref{sec: Multiclass Classification: Mapping The Boundaries Of Amplitude Encoding} shares this configuration except where explicitly noted there.

\begin{table}[H]
	\centering
	\caption{Consolidated experimental configuration for the proposed VQC pipeline.}
	\label{tab: config}
	\begin{tabular}{@{}ll@{}}
		\toprule
		\textbf{Component} & \textbf{Setting} \\
		\midrule
		Dataset & NSL-KDD, official 20\% subset (\texttt{KDDTrain+\_20Percent}) \\
		Records & 25,192 (80/20 stratified train/test: 20,155 / 5,037) \\
		Categorical encoding & Target Encoding (smoothed) \\
		Scaling & StandardScaler \\
		Dimensionality reduction & PCA, 16 components ($\sim$86.78\% explained variance) \\
		State preparation & Amplitude encoding, 4 qubits ($2^4 = 16$) \\
		Trainable ansatz & \texttt{RealAmplitudes}, \texttt{reps=5} (24 parameters) \\
		Measurement & Z-basis with parity label mapping \\
		Optimizer & COBYLA (gradient-free), max.\ 500 iterations \\
		Loss monitored & Cross-Entropy (via custom callback) \\
		Simulator & Classical statevector simulation \\
		\bottomrule
	\end{tabular}
\end{table}

The primary optimization of the 24 quantum weights was driven by the COBYLA algorithm. Operating strictly as a gradient-free optimizer \cite{BonetMonroig2023}, it mitigates the massive computational overhead associated with analytic quantum derivatives. The optimization was constrained to a maximum of 500 iterations, actively guided by a custom callback function monitoring the Cross-Entropy Loss to track objective convergence.

The definitive evaluation of the fully trained quantum model was executed strictly on the isolated Test Set (5,037 network flows). As detailed in Table \ref{tab1}, the VQC established a robust empirical baseline, generalizing to unseen data with an overall accuracy of 88\%, supported by highly balanced Precision and Recall metrics across both normal and malicious classes.

\begin{table}[H]
	\centering
	\caption{Binary classification report for the proposed 4-qubit VQC.}
	\label{tab1}
	\begin{tabular}{@{}lccc@{}}
		\toprule
		\textbf{Metric} & \textbf{Normal (0)} & \textbf{Attack (1)} & \textbf{Accuracy} \\
		\midrule
		Precision & 0.90 & 0.86 & - \\
		Recall & 0.87 & 0.89 & - \\
		F1-Score & 0.89 & 0.88 & 0.88 (88\%) \\
		\bottomrule
	\end{tabular}
\end{table}
\subsection{Methodological safeguard: isolating the dimensionality constraint} \label{subsec: Methodological Safeguard: Isolating the Dimensionality Constraint}

While an 88\% accuracy demonstrates functional quantum learning, it introduces a critical research dichotomy: Is this performance ceiling an artifact of insufficient optimization (the optimizer collapsing into a shallow local minimum), or is it an absolute representational bound dictated by the quantum architecture itself?

To resolve this, the methodological safeguard of concurrent training via the stochastic SPSA optimizer was triggered. Specifically engineered to bypass Barren Plateaus through simultaneous random parameter perturbations \cite{Kandala2017, Kulshrestha2022}, SPSA actively tests the global loss landscape. The parallel evaluation conclusively demonstrated that both fundamentally distinct algorithms plateaued at exactly the same accuracy limits. This behavioral symmetry provides definitive empirical proof that the 88\% ceiling is not an optimization failure. Rather, it is a structural consequence of Amplitude Encoding: forcefully compressing 16 independent data variance vectors into the overlapping probability amplitudes of just 4 qubits inherently restricts the model's geometric flexibility, permanently capping its linear separability. This aligns directly with recent comparative studies in network intrusion detection, which confirm that highly compressed quantum embeddings severely restrict classification expressibility \cite{Kadi2025}.

\subsection{Classical benchmarking \& the dimensionality bottleneck} \label{subsec: Classical Benchmarking & the Dimensionality Bottleneck}

Having mapped the intrinsic expressive limits of the parameterized quantum circuit, the VQC was subjected to a rigorous benchmark against conventional classical architectures possessing strictly equivalent or simplified structural complexity, fed with the exact same input representation (16 PCA components).

The linear baseline was established using Logistic Regression \cite{Hastie2009}. Operating as the simplest possible classifier (assigning a single linear weight to each feature plus a bias, totaling 17 parameters—fewer than the VQC's 24), it achieved 92\% accuracy in milliseconds. This outcome definitively proves that the highly compressed PCA-16 dataset inherently retains linear separability, explicitly highlighting the information loss occurring exclusively during the quantum amplitude overcompression.

To perfectly isolate the effect of the 4-qubit dimensionality restriction, a highly constrained classical neural network (a ``Tiny MLP'') was designed. To structurally mirror the quantum state space bottleneck, the hidden layer was strictly limited to exactly 4 nodes. The total network complexity equated to 73 parameters, optimized via the Adam solver for 1,000 epochs. Crucially, while the quantum circuit remains bound by linear unitary matrices, the classical MLP leverages the Rectified Linear Unit (ReLU) activation function \cite{Nair2010} to warp the decision space non-linearly.

\begin{figure}[H]
	\centering
	\begin{tikzpicture}[scale=1.15, every node/.style={scale = 1.100}]
		\definecolor{cMax}{HTML}{3f007d}
		\definecolor{cHigh}{HTML}{54278f}
		\definecolor{cLow}{HTML}{f2f0f7}
		\definecolor{textDark}{HTML}{3f007d}
		\fill[cMax] (0,2) rectangle (2,4); \node[text=white] at (1,3) {2613};
		\fill[cLow] (2,2) rectangle (4,4); \node[text=textDark] at (3,3) {77};
		\fill[cLow] (0,0) rectangle (2,2); \node[text=textDark] at (1,1) {78};
		\fill[cHigh] (2,0) rectangle (4,2); \node[text=white] at (3,1) {2269};
		\draw[black, thin, step=2] (0,0) grid (4,4);
		\draw[black, thick] (0,0) rectangle (4,4);
		\node[left, xshift=-0.1cm] at (0,3) {Normal};
		\node[left, xshift=-0.1cm] at (0,1) {Attack};
		\node[below, yshift=-0.1cm] at (1,0) {Normal};
		\node[below, yshift=-0.1cm] at (3,0) {Attack};
		\node[rotate=90] at (-1.9, 2) {True label};
		\node at (2, -0.9) {Predicted label};
		\shade[top color=cMax, bottom color=white] (4.5,0) rectangle (4.8,4);
		\draw[black, thin] (4.5,0) rectangle (4.8,4);
		\foreach \y/\val in {0/500, 0.8/1000, 1.6/1500, 2.4/2000, 3.2/2500} {
			\draw (4.8, \y) -- (4.9, \y) node[right, font=\scriptsize] {\val};
		}
	\end{tikzpicture}
	\caption{Performance evaluation of the classical baseline: Confusion Matrix of the 4-Node Tiny MLP, simulating the extreme dimensional constraint of the quantum architecture.}
	\label{fig: tmlp}
\end{figure}
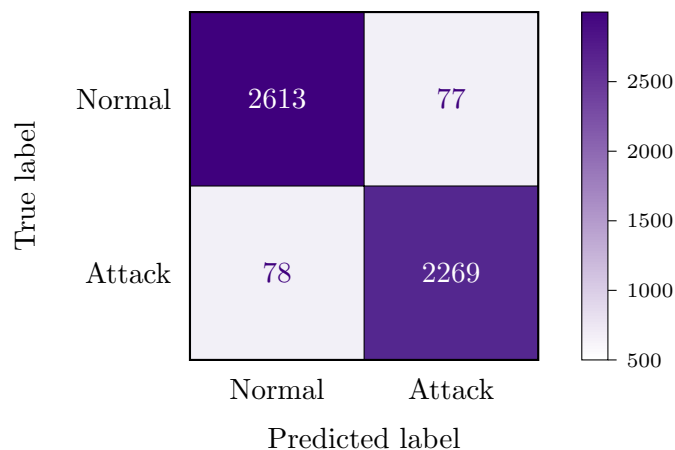

As visually demonstrated by the confusion matrix in Figure \ref{fig: tmlp}, this 4-node architectural analog achieved an exceptional overall accuracy of 97\%, vastly outperforming the VQC. This aligns perfectly with recent comparative literature confirming that classical neural networks currently scale and generalize more effectively than parameterized quantum models due to the severe hardware constraints of the NISQ era \cite{Eze2025}. Furthermore, this performance degradation under NISQ conditions is a well-documented phenomenon across domains; for instance, recent evaluations in speech emotion recognition similarly report classical baselines severely outperforming hybrid quantum pipelines \cite{Norval2025}.

These comparative results establish a highly constructive paradigm for anomaly detection: classical algorithms can successfully interpret aggressively reduced dimensions provided they possess non-linear expressiveness. Conversely, the 4-qubit VQC structurally fails to disentangle these tightly packed probability amplitudes using only linear unitary transformations. Therefore, realizing true quantum advantage demands abandoning aggressive amplitude compression in favor of broader representation topologies. This conclusion is explicitly corroborated by recent comparative analyses (2026), which definitively establish that Angle Encoding significantly outperforms Amplitude Encoding in capturing complex non-linear relationships for medical and diagnostic classifications \cite{Hossain2026}.

\section{Multiclass classification: mapping the boundaries of Amplitude Encoding} \label{sec: Multiclass Classification: Mapping The Boundaries Of Amplitude Encoding}

While in the binary classification stage (Section 7) the model successfully demonstrated an 88\% capability to distinguish malicious from normal traffic, a critical test for any Intrusion Detection System (IDS) is the granular categorization of threat typologies. To probe the expressive limits of the VQC under this configuration, a multi-class classification experiment was first deployed across the four grouped attack families (DoS, Probe, R2L, U2R; Section \ref{subsec: The Accuracy Paradox & Imbalance Sensitivity}), after which the experiment was repeated at the finer granularity of the 22 raw NSL-KDD attack categories to test whether the observed bottleneck persists independently of label grouping (Section \ref{subsec: Data Balancing & Degenerate Mode Collapse}).

\subsection{The accuracy paradox \& imbalance sensitivity} \label{subsec: The Accuracy Paradox & Imbalance Sensitivity}

During the initial training on the natively imbalanced dataset, a severe statistical skew was present (e.g., thousands of DoS entries versus merely 10 U2R entries). While the initial evaluation yielded an ostensibly high overall accuracy of 91\%, the granular classification report (Table \ref{tab2}) exposed a critical failure in class-specific performance.

\begin{table}[H]
	\centering
	\caption{Classification Report (Imbalanced Multiclass VQC).}
	\label{tab2}
	\begin{tabular}{@{}lccc@{}}
		\toprule
		\textbf{Class} & \textbf{Precision} & \textbf{Recall} & \textbf{F1-Score} \\
		\midrule
		DoS (0) & 0.99 & 0.91 & 0.95 \\
		Probe (1) & 0.67 & 0.96 & 0.79 \\
		R2L (2) & 0.00 & 0.00 & 0.00 \\
		U2R (3) & 0.00 & 0.00 & 0.00 \\
		\bottomrule
	\end{tabular}
\end{table}

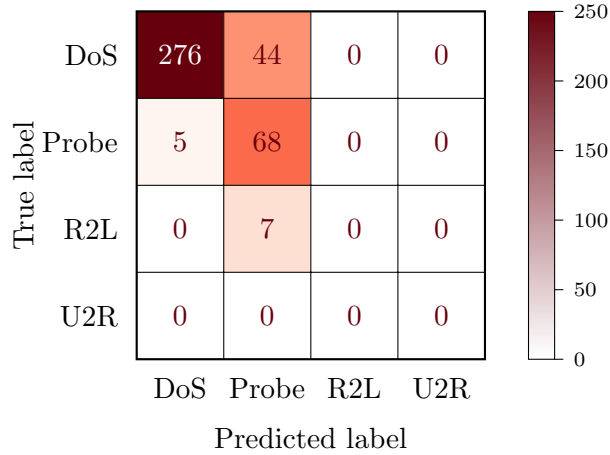
\begin{figure}[H]
	\centering
	\begin{tikzpicture}[scale=1.15, every node/.style={scale=1.100}]
		\definecolor{c276}{HTML}{67000d}
		\definecolor{c68}{HTML}{fb6a4a}
		\definecolor{c44}{HTML}{fc9272}
		\definecolor{c7}{HTML}{fee0d2}
		\definecolor{c5}{HTML}{fff5f0}
		\definecolor{c0}{HTML}{ffffff}
		\definecolor{textDark}{HTML}{67000d}
		\fill[c276] (0,3) rectangle (1,4); \node[text=white] at (0.5,3.5) {276};
		\fill[c44] (1,3) rectangle (2,4); \node[text=textDark] at (1.5,3.5) {44};
		\fill[c0] (2,3) rectangle (3,4); \node[text=textDark] at (2.5,3.5) {0};
		\fill[c0] (3,3) rectangle (4,4); \node[text=textDark] at (3.5,3.5) {0};
		\fill[c5] (0,2) rectangle (1,3); \node[text=textDark] at (0.5,2.5) {5};
		\fill[c68] (1,2) rectangle (2,3); \node[text=textDark] at (1.5,2.5) {68};
		\fill[c0] (2,2) rectangle (3,3); \node[text=textDark] at (2.5,2.5) {0};
		\fill[c0] (3,2) rectangle (4,3); \node[text=textDark] at (3.5,2.5) {0};
		\fill[c0] (0,1) rectangle (1,2); \node[text=textDark] at (0.5,1.5) {0};
		\fill[c7] (1,1) rectangle (2,2); \node[text=textDark] at (1.5,1.5) {7};
		\fill[c0] (2,1) rectangle (3,2); \node[text=textDark] at (2.5,1.5) {0};
		\fill[c0] (3,1) rectangle (4,2); \node[text=textDark] at (3.5,1.5) {0};
		\fill[c0] (0,0) rectangle (1,1); \node[text=textDark] at (0.5,0.5) {0};
		\fill[c0] (1,0) rectangle (2,1); \node[text=textDark] at (1.5,0.5) {0};
		\fill[c0] (2,0) rectangle (3,1); \node[text=textDark] at (2.5,0.5) {0};
		\fill[c0] (3,0) rectangle (4,1); \node[text=textDark] at (3.5,0.5) {0};
		\draw[black, thin] (0,0) grid (4,4);
		\draw[black, thick] (0,0) rectangle (4,4);
		\node[left, xshift=-0.1cm] at (0,3.5) {DoS};
		\node[left, xshift=-0.1cm] at (0,2.5) {Probe};
		\node[left, xshift=-0.1cm] at (0,1.5) {R2L};
		\node[left, xshift=-0.1cm] at (0,0.5) {U2R};
		\node[below, yshift=-0.1cm] at (0.5,0) {DoS};
		\node[below, yshift=-0.1cm] at (1.5,0) {Probe};
		\node[below, yshift=-0.1cm] at (2.5,0) {R2L};
		\node[below, yshift=-0.1cm] at (3.5,0) {U2R};
		\node[rotate=90] at (-1.3, 2) {True label};
		\node at (2, -0.9) {Predicted label};
		\shade[top color=c276, bottom color=white] (4.5,0) rectangle (4.8,4);
		\draw[black, thin] (4.5,0) rectangle (4.8,4);
		\foreach \y/\val in {0/0, 0.8/50, 1.6/100, 2.4/150, 3.2/200, 4.0/250} {
			\draw (4.8, \y) -- (4.9, \y) node[right, font=\scriptsize] {\val};
		}
	\end{tikzpicture}
	\caption{Confusion matrix for the imbalanced four-class VQC experiment, visually demonstrating the model's complete failure to predict minority classes (R2L and U2R).}
	\label{fig: imbalanced_cm}
\end{figure}

This scenario represents a textbook example of the accuracy paradox \cite{Luque2019}. Driven by the dominance of DoS attacks, the COBYLA optimizer converged to a majority-class heuristic to mathematically minimize the loss function. This vulnerability of Quantum Machine Learning models to severe class imbalance is a well-documented challenge in recent literature \cite{Kwon2025}. As visually confirmed by the confusion matrix in Figure \ref{fig: imbalanced_cm}, the model almost exclusively predicted DoS and Probe categories, leaving the lower-frequency intrusion vectors (R2L and U2R) with a recall of exactly 0\%.

\subsection{Data balancing \& degenerate mode collapse} \label{subsec: Data Balancing & Degenerate Mode Collapse}

To definitively determine whether the failure documented in Section \ref{subsec: The Accuracy Paradox & Imbalance Sensitivity} was merely a byproduct of classical statistical distribution or a structural limitation of the quantum architecture itself, a class-balanced training set was synthetically constructed at the finer granularity of the 22 raw NSL-KDD attack categories (exactly 200 samples per class, 4{,}400 samples total), via random oversampling with replacement in every minority subset \cite{Buda2018}, forcing the circuit to confront every individual attack signature with equal mathematical weight.

The trained model was then evaluated \textit{in-sample}---that is, on the very same balanced training data it had just been fit on, rather than on a held-out test partition. We use this protocol deliberately, but as a \emph{diagnostic probe} rather than a benchmark: a model that cannot fit data it has already observed during training is unlikely to be limited by overfitting or distributional shift, which points towards a representational rather than a generalization difficulty. We stress, however, that in-sample failure alone does not conclusively isolate the encoding as the cause---an insufficiently expressive ansatz, a limited optimizer budget, or the measurement-to-label decoding could each produce a similar effect. We therefore treat the result below as evidence \emph{consistent with} a capacity bottleneck, and we revisit these competing explanations explicitly in subsection \ref{subsec: Limitations, Threats To Validity & Future Directions}.

The resulting behavior provided a clear and mathematically informative answer: the model's predictive capacity deteriorated to an overall in-sample accuracy of just 9\%. With 22 classes, a uniform random-guessing baseline would be expected to yield approximately 4.5\% accuracy; the model's 9\% is therefore numerically above this baseline, but the classification report reveals that this is not evidence of partial success. Of the 22 classes, only two ('back' and 'buffer\_overflow') received any non-zero predictions at all, each with a recall of 1.00 but a precision of only approximately 0.10---meaning the model funneled the overwhelming majority of all 4{,}400 training samples, regardless of true class, into just these two output bins. The remaining 20 classes, including 'normal' itself, received a precision, recall, and F1-score of exactly 0.00. This is not the signature of noisy or near-random classification; it is the signature of a degenerate mode collapse, in which the optimizer converged to a fixed point that ignores the input entirely and outputs from a tiny subset of the available classes irrespective of the data presented to it.

This pattern is consistent with the mechanism proposed for the coarser 4-class case in Section \ref{subsec: The Accuracy Paradox & Imbalance Sensitivity}: because amplitude encoding forces 16 distinct, non-linear classical features into overlapping probability amplitudes within just 4 qubits, the resulting quantum states representing different attack families exhibit extremely high fidelity (similarity) to one another. Unable to mathematically disentangle these highly overlapping states using linear unitary operations, the ansatz converges to a degenerate local minimum in which only the few classes occupying the most separable regions of the compressed Hilbert space receive any non-trivial measurement signal. Consequently, the measurement expectation values collapse systematically onto a small handful of class eigenvalues rather than spreading across the full label set, resulting in the systemic, non-random misclassification pattern documented above.

\section{Discussion \& Conclusions} \label{sec: Discussion & Conclusions}

This research successfully mapped the structural limits of quantum computing against a rigorous cybersecurity benchmark (NSL-KDD). The central empirical takeaway establishes a clear directive for the field: extreme feature compression via amplitude encoding is practically inadequate for the granular detection of complex, multi-class network intrusions in the NISQ era.

\subsection{Charting the limits of the NISQ bottleneck} \label{subsec: Charting The Limits of the NISQ Bottleneck}

While ``Quantum Economy'' (embedding 16 features into 4 qubits) initially appears as a viable strategy for navigating the stringent hardware constraints of NISQ systems, our experiments indicate clear practical limits for highly nonlinear problems under this configuration. Although the architecture bounded a binary classification threshold at 88\% accuracy on the held-out partition, introducing the complexity of multi-class attacks surfaced the system's breaking point, with a degenerate mode collapse yielding a mere 9\% in-sample accuracy on the 22-class balanced attack set. We frame this as a configuration-level observation---scoped to amplitude encoding with a shallow ansatz and gradient-free optimization on the 20\% subset---rather than as a universal bound on quantum classification.

To guarantee the scientific validity of these findings, our methodology systematically isolated the core architectural variables. First, the optimization invariance demonstrated by the identical performance plateaus of both the deterministic COBYLA and the noise-resilient SPSA proved that the stagnation is an immutable boundary of the quantum architecture itself, rather than an artifact of localized training failures. Second, the baseline imperative was established by subjecting a classical 4-node Tiny MLP to identical compression algorithms, which yielded a superior 97\% accuracy. This successfully demonstrates that the dataset's intrinsic value is preserved post-compression, provided it is processed via non-linear transformations (such as ReLU)—computations that current parameterized quantum circuits, bound strictly by linear unitary matrices, cannot mathematically emulate.

\subsection{The roadmap to Quantum Advantage: angle-entangled 16-qubit VQC} \label{subsec: The Roadmap To Quantum Advantage: Angle-Entangled 16-Qubit VQC}

Recognizing these structural limitations fundamentally validates the transition to new Quantum Machine Learning models. To definitively bypass the ``Quantum Bottleneck,'' an uncompressed architecture is required. We propose a structure utilizing a direct ``One Feature per Qubit'' approach via Angle Encoding \cite{PerezSalinas2020}.

By mapping each feature to an independent, dedicated qubit utilizing a unitary rotation gate $R_y(x_i)$, Cover’s Theorem on pattern separability \cite{Samuelson2011} is directly leveraged. As established by Cover, projecting non-linear data into a sufficiently high-dimensional space significantly increases the probability that the data points will become linearly separable. Projecting the feature vector into an exponentially expanded quantum Hilbert space ($2^{ 16 } = 65,536$ dimensions) mathematically guarantees that complex, entangled attack vectors are mapped into true linear separability.

Furthermore, recent fundamental evaluations of quantum perceptrons confirm that quantum models inherently possess a higher storage capacity (Gardner volume) than classical perceptrons, effectively functioning as non-monotonic activation functions that expand the viable classification space \cite{Urushibata2025}. This theoretical capacity can be practically unlocked by leveraging the expanded effective dimension of quantum neural networks \cite{Abbas2021}, ensuring superior generalization bounds compared to classical counterparts \cite{HuangHY2021}. Within this expansive space, network correlations can be captured using a fully cyclically entangled layer (Ring-Topology CNOTs). This topology encodes packet relationships as a fundamental physical property of the quantum state, potentially eliminating the requirement for the massive classical weight matrices demanded by traditional deep learning networks.

\subsection{Limitations, threats to validity \& future directions} \label{subsec: Limitations, Threats To Validity & Future Directions}

Several boundaries constrain the scope of the claims made in this paper, and we state them explicitly so that our results are read as scoped empirical observations rather than universal bounds.

First, our central observation---the in-sample mode collapse on the 22-class balanced set---does not by itself isolate amplitude encoding as the sole cause. The experiment simultaneously fixes the encoding, a shallow \texttt{RealAmplitudes} ansatz, a gradient-free optimizer with a finite iteration budget, and a fixed measurement-to-label decoding scheme. Any of these could contribute to the observed collapse, and a targeted ablation that varies ansatz depth and optimizer budget while holding the encoding fixed would be required to attribute the effect to the encoding conclusively. We treat the encoding as the most plausible contributor---in line with the theoretical expressibility limits reported by Kadi et al. \cite{Kadi2025}---but we do not claim to have isolated it.

Second, all experiments are conducted on held-out partitions of the official 20\% NSL-KDD research subset (\texttt{KDDTrain+\_20Percent}); the separate \texttt{KDDTest+} evaluation file, which contains attack types absent from training, was not used, owing to the simulation-cost constraints described in Section \ref{sec: NSL-KDD Dataset Analysis}. Our reported accuracies therefore characterize behavior within this subset and should not be interpreted as estimates of generalization to the full benchmark. Evaluation on \texttt{KDDTest+} is a priority for future work. The reported figures correspond to representative training runs; given the known variance of variational quantum training, a multi-seed statistical characterization is a natural extension.

Third, on the computational side, empirical simulations of the expanded 16-qubit architecture mapped a clear computational boundary, exceeding 10 hours of processing for merely 50 records during initial trials. Resource profiling isolated this to a lack of parallelizability: the sequential Markov chains dictating gradient-free optimizers, compounded by the Python Global Interpreter Lock (GIL), resulted in unsustainable CPU overhead.

To render the Angle-Entangled model computationally viable and advance the field beyond its current exploratory phase \cite{Wichert2024}, future research must transition to high-performance execution strategies. This includes migrating to dynamic, low-level compilation ecosystems such as JAX or PennyLane, where techniques like adjoint differentiation will enable fast, analytical gradient computations, drastically reducing the exponential overhead of parameter-shift rules \cite{Bergholm2022}. Furthermore, bypassing Python entirely by authoring native execution scripts in C/C++ will eliminate the GIL bottleneck and maximize CPU utilization. Ultimately, realizing the full potential of this architecture requires execution on physical Quantum Processing Units (QPUs), removing the simulation constraints inherent to the NISQ era and allowing quantum anomaly detection to be evaluated in real-time network environments.

\bibliographystyle{ieeetr}
\bibliography{EVQCCD}

\end{document}